\documentclass[preprint,12pt]{elsarticle}

\usepackage{amssymb}
\usepackage{amsmath}
\usepackage[colorlinks=true, allcolors=blue, hidelinks, breaklinks]{hyperref}

\usepackage{framed,enumitem}

\usepackage[dvipsnames]{xcolor}
\usepackage{listings}
\lstdefinestyle{mystyle}{
    keywordstyle=\bfseries,
    basicstyle=\ttfamily\footnotesize,
    breakatwhitespace=false,
    breaklines=true,
    keepspaces=true,
    showspaces=false,
    showstringspaces=false,
    showtabs=false,
    tabsize=2,
    frame=lines
}
\colorlet{punct}{red!80!black}
\definecolor{delim}{RGB}{20,105,176}
\colorlet{numb}{magenta!80!black}

\lstdefinelanguage{grammar}
{
    morekeywords={country, location, country, fragment, endDate, teamSize, iterations, teamParticipants, age, workplaceType, educationalLevel, spokenLanguages, skillLevel, tenure, participantId, description, startDate, reportingMethod, reportingPlatform, bodies, useCases, participants, release, targetCommunities, organizationsAndTargetCommunities, governances, socialContexts, adaptations, name, socioEconomicStatus, startDate }
    morecomment=[l]
    morestring=[b]"
    columns=flexible,
    keepspaces=true,
    showstringspaces=false,
    basicstyle=\ttfamily\footnotesize,
    commentstyle=\color{green},
    keywordstyle=\color{RoyalPurple},
    stringstyle=\color{gray},
    morestring=[b]",
    morestring=[b],
    xleftmargin=1ex,
    numbers=left,
    numberstyle=\scriptsize,
    stepnumber=1,
    numbersep=3pt,
    moredelim=[s][\color{gray}]{'}{' }
}

\lstdefinelanguage{questionnaire}
{
    morekeywords={Section},
    keepspaces=true,
    showstringspaces=false,
    basicstyle=\ttfamily\footnotesize,
    keywordstyle=\color{RoyalPurple},
}

\lstdefinelanguage{DSL}
{
  morekeywords={targetCommunity, targetCommunities, description, ageRange, locations, workplaceType, presidential, countries, Spain, educationalLevels, shortCycleTertiary, primary, earlyChildhood, spokenLanguages, socioEconomicStati, lowerClass, lowerMiddleClass, skillLevels, beginner, adaptation, relatedTeams, developmentTeam, startDate, teamSize, ethnicities, genders, religiousBeliefs, averageTenure},
  morecomment=[l]{Purpose, Tasks, Gaps, HealthCare, Images, skinImages, Numerical, age,},
  morestring=[b]",
   columns=flexible,
    keepspaces=true,
    showstringspaces=false,
    basicstyle=\ttfamily\footnotesize,
    commentstyle=\color{green},
    keywordstyle=\color{purple},
    stringstyle=\color{gray},
    xleftmargin=1ex,
    numbers=left,
    numberstyle=\scriptsize,
    stepnumber=1,
    numbersep=3pt,
    literate=
     *
      {Categorical-Distribution}{{{\color{purple}{Categorical Distribution}}}}{1}
      {DataInstances}{{{\color{purple}{Data Instances}}}}{1}
      {Record-Data}{{{\color{black}{Record-data}}}}{1}
      {:}{{{\color{delim}{:}}}}{1}
      {,}{{{\color{punct}{,}}}}{1}
      {\{}{{{\color{delim}{\{}}}}{1}
      {\}}{{{\color{delim}{\}}}}}{1}
      {[}{{{\color{delim}{[}}}}{1}
      {]}{{{\color{delim}{]}}}}{1}
}
\usepackage[dvipsnames]{xcolor}
\newcommand\change[1]{#1}
\newcommand\changetwo[1]{}

\usepackage{graphicx}
\usepackage{multirow}
\usepackage{booktabs} 
\usepackage{geometry}
\usepackage{tabularx} 
\usepackage{adjustbox}

\renewcommand{\arraystretch}{2}

\definecolor{codegreen}{rgb}{0,0.6,0} 
\definecolor{codegray}{rgb}{0.5,0.5,0.5}
\definecolor{codepurple}{rgb}{0.58,0,0.82} 
\definecolor{backcolour}{rgb}{1,1,1}
\lstdefinestyle{codeStyle}{
    backgroundcolor=\color{backcolour},   
    commentstyle=\color{codegreen},
    keywordstyle=\color{magenta},
    numberstyle=\tiny\color{codegray},
    stringstyle=\color{codepurple},
    basicstyle=\ttfamily\footnotesize,
    breakatwhitespace=false,         
    breaklines=true,                 
    captionpos=b,                    
    keepspaces=true,                 
    numbers=none,                    
    numbersep=5pt,                  
    showspaces=false,                
    showstringspaces=false,
    showtabs=false,  
    frame=single,
    tabsize=2,
}

\usepackage{array} 
\newcommand{\PreserveBackslash}[1]{\let\temp=\\#1\let\\=\temp}
\newcolumntype{C}[1]{>{\PreserveBackslash\centering}p{#1}}
\newcolumntype{R}[1]{>{\PreserveBackslash\raggedleft}p{#1}}
\newcolumntype{L}[1]{>{\PreserveBackslash\raggedright}p{#1}}

\begin{document}

\begin{frontmatter}



\title{The Software Diversity Card: A Framework for Reporting Diversity in Software Projects}


\author[BSC]{Joan Giner-Miguelez}

\author[UOC]{Sergio Morales}
\author[UOC]{Sergio Cobos}
\author[UOC]{Javier Luis Cánovas Izquierdo}
\author[UOC]{Robert Clarisó}
\author[LIST,UL]{Jordi Cabot}

\affiliation[BSC]{organization={Barcelona Supercomputing Center, BSC-CNS},
             city={Barcelona},
            state={Spain},
            country={}}

\affiliation[UOC]{organization={Universitat Oberta de Catalunya},
            city={Barcelona},
            state={Spain},
            country={}}

\affiliation[LIST]{organization={Luxembourg Institute of Science and Technology},
            city={Esch-sur-Alzette},
            state={Luxembourg},
            country={}}

\affiliation[UL]{organization={University of Luxembourg},
            city={Esch-sur-Alzette},
            state={Luxembourg},
            country={}}

\begin{abstract}

\textbf{Context:} Interest in diversity in software development has significantly increased in recent years. Reporting on diversity in software projects can enhance user trust and assist regulators in evaluating adoption. Recent AI directives include clauses that mandate diversity information during development, highlighting the growing interest of public regulators. However, current documentation often neglects diversity in favor of technical features, partly due to a lack of tools for its description and annotation. 

\textbf{Objectives:} This work introduces the Software Diversity Card, a structured approach to document and share diversity-related aspects within software projects. It aims to profile the various teams involved in software development and governance, including user groups in testing and software adaptations for diverse social groups.
 
\textbf{Methods:} We performed a literature review on diversity and inclusion in software development and an analysis of 1,000 top-starred Open Source Software (OSS) repositories in \textsc{GitHub} to identify diversity-related information. \change{Moreover, we present a diversity modeling language, a toolkit for generating the cards using that language, and a study of its application in two real-world software projects.}

\textbf{Results:} Despite the growing awareness of diversity in the research community, our analysis found a notable lack of diversity reporting in OSS projects. Applying the card to real-world examples highlighted challenges like balancing anonymity with transparency, managing sensitive data, and ensuring authenticity.

\textbf{Conclusion:} We believe that our proposal can enhance diversity practices in software development, 
support public administrations in software assessment, and help businesses promote diversity as a key asset.

\end{abstract}




\end{frontmatter}



\section{Introduction}

The diversity aspects around software projects have been a growing concern in the last years~\cite{bjorn2023diversity,albusays2021diversity}. Recent research has proven, among others, that diverse and inclusive teams can produce better software \cite{yang2022gender}, benefit open-source \mbox{communities \cite{vasilescu2015gender}}, and facilitate understanding of the human factor involved in the \mbox{development process \cite{GUNATILAKE2024107489}}. As a result, many efforts have been made to create diverse and inclusive software development \mbox{teams \cite{hyrynsalmi2024making,oliveira2024navigating, zhao2024early}.} However, the impact of diversity is not limited to development teams; it has also been shown to benefit software end users. Assessing the diversity of testers or user feedback may be critical for software projects to meet the needs of their target users \cite{wang2022context}.

Learning about the people engaged in software product creation can help users better understand the software they use, \change{ thereby increasing trust \cite{fogg2003users,steinmacher2015systematic}}. For example, a period tracker app designed for women may inspire greater user confidence if the developers and testers include women sharing similar experiences and needs. Similarly, when developing software for specific social groups, it is essential to consider their context and unique characteristics. For example, Bangladeshi fisherfolk faced significant barriers in adopting new software technologies due to factors such as low digital literacy, varying language styles, and their community's traditional knowledge transmission methods. These kinds of barriers can be mitigated by gaining a better understanding of the social group's context and specific \change{needs \cite{kanij2023developing}}.  

Concerns about sharing diversity information are also increasing in other domains, impacting public regulatory bodies worldwide. For instance, influential documentation frameworks in the Machine Learning (ML) field stress the need to document the diversity of actors involved in the data creation processes \cite{gebru2021datasheets, pushkarna2022data, akhtar2024croissant}. These recommendations have been incorporated into recent regulations on artificial intelligence, such as the European AI Act\footnote{EU AI Act required documentation - Annex IV: \url{ https://www.euaiact.com/annex/4}} and the US AI Bill of Rights\footnote{US AI Bill of Rights: \url{https://bidenwhitehouse.archives.gov/ostp/ai-bill-of-rights}}, and are expected to become a fundamental practice in developing future responsible AI applications.

This increasing interest in transparency and diversity illustrates the utility of this information for public regulators. For example, regulations requiring gender balance in software development can encourage more women to enter the field, or policymakers might promote economic growth in particular areas by requiring developers to come from specific regions. Furthermore, similar to the Bangladeshi fisherfolk's example, an administration might ensure that the app's beta tests \cite{leicht2017leveraging} are carried out across a specific population to ensure that the app is properly designed for them. \looseness=-2

Despite the clear benefits of reporting diversity-related aspects of software projects, there are currently no standardized documentation practices or concrete reporting techniques for this purpose. \change{Existing documentation in software engineering primarily emphasizes technical features, often overlooking social and diversity considerations, while current machine learning documentation frameworks tend to focus on data assets, frequently neglecting the broader complexities of software projects.}

To fill this gap, and inspired by recent ML data documentation frameworks, we propose the Software Diversity Card as a comprehensive approach to documenting and reporting diversity information in software projects. The card allows the profiling of the various types of teams involved in the development process, including the development team, non-coding contributors to the project \cite{canovas2022analysis} (\emph{e.g.}, issue reporters, translators), final users involved in testing (\emph{e.g.}, crowd testers \cite{leicht2017leveraging}), and those involved in the software governance. Also, the card aims to capture the software usage context by profiling the intended target users and the software adaptations for specific social groups, such as accessibility features to particular disabilities \cite{bi2022accessibility}. 

Our reporting framework could benefit the open-source (OS) community diversity practices by being integrated along with other project documentation, such as the contributing guidelines \cite{elazhary2019not}, the codes of conduct \cite{tourani2017code} or governance files~\cite{izquierdo2023governance} in public repositories. It may also benefit the research community by being included in the submission guidelines for scientific journals that publish original software publications~\cite{smith2016software}. Furthermore, our framework can assist businesses in viewing diversity as a valuable asset and help public administrations by incorporating diversity into their software assessment criteria.

In the remainder of this document, Section~\ref{sec:related} presents the background and related work on diversity and human aspects in software development, while Section~\ref{empirical} empirically evaluates the current reporting diversity practices in open-source communities motivating the need for our proposal. Then, Section~\ref{sec:card} presents the main concepts and structure of the card. Section~\ref{sec:model} describes the development of the proposal, introducing a \change{diversity} domain-specific language and a toolkit (including a language plugin\footnote{Language plugin repository: \url{https://github.com/SOM-Research/SoftwareDiversityCard}} and a form-based web editor\footnote{Online demo of the web editor: \url{https://huggingface.co/spaces/JoanGiner/SoftwareDiversityCard-Generator}}) that supports developers in creating diversity cards for their projects. In Section~\ref{sec:usecases}, we present two running examples of real-world software projects described using our proposal. Finally, in Section~\ref{sec:challenges}, we discuss the challenges and limits of our study, while in Section~\ref{sec:conclusions}, we present the conclusion and open research lines on reporting diversity in software projects.

\section{State of the art}
\label{sec:related}

This section reviews prior research on diversity and inclusion in software development. \change{We begin by examining the dimensions that previous studies have used to characterize diversity, which provides the foundation for our own characterization of teams in software projects}. Next, we explore research focused on diversity within different types of teams in software projects, including both development teams and non-coding contributors. We then discuss the significance of diversity for end users and how projects adapt to meet the needs of specific populations. Finally, we provide a brief overview of emerging machine learning documentation frameworks, \change{highlighting their opportunities and limitations in software projects.}

\change{\subsection{Characterizing diversity in software}}
\label{sec:related:character}

\change{Relevant works on Software Developers’ Diversity and Inclusion (SDDI) have shown that economic and political structures shape who gains access to technology, whose perspectives are prioritized in software development, and whose biases become embedded in digital systems \cite{bjorn2023diversity, albusays2021diversity}. These findings highlight the importance of examining the profiles of those involved in the complex ecosystem of software projects. A central contribution of prior work has been the identification of a set of dimensions for studying and characterizing diversity within software engineering. }

\change{In this regard, Dutta et al. \cite{dutta2023diversity} analyzed 79 research papers and 105 participant studies in the SDDI community, identifying the most common dimensions used to describe individuals involved in software projects. These include gender, age, ethnicity/race, socioeconomic status, and sexual orientation, as well as professional attributes such as tenure, job title, language expertise, and education level. The selection of these dimensions is also reflected in recent influential studies \cite{bjorn2023diversity} and community-driven initiatives, such as special issues on the diversity crisis in software development \cite{albusays2021diversity}, that underscored the need to to account also for disabilities and intersectionality during the study of diversity.}  

\change{Together, these efforts establish a baseline for characterizing diversity in software projects. While this characterization is dynamic and may evolve as new societal needs emerge, it provides the foundation for our selection of dimensions in this study, shown in Figure \ref{fig:mm_entities_individuals}. To underscore the relevance of these dimensions, the following section reviews prior work that has examined diversity across them, with a focus on both development teams first and then on other types of teams involved directly and indirectly in software projects.}

\subsection{Diversity on software development teams}
\label{sec:related:developers}

\change{The study of the composition of open-source projects has been a focus of interest of the research community. One of the seminal studies is the proposal of the onion model \cite{crowston2005social}, which studies the software project’s social structure as concentric layers of contributors. The model highlights core contributors at the center, surrounded by peripheral and occasional participants, illustrating roles, responsibilities, and patterns of collaboration within the community. An excellent case study illustrating the structure of real-world open-source software projects is the GNOME project \cite{german2003gnome} and subsequent studies on the composition and inherent nature of the contributors \cite{oreg2008exploring,  robles2014floss}. Going beyond the composition, posterior studies \cite{santos2013attraction} explored how both intrinsic and extrinsic factors shape contributor engagement, relating to the focus of interest of this work.}

The composition of the development team has also been studied regarding the gender dimensions. \change{Studies show that the lack of gender diversity remains an issue} \cite{bosu2019diversity}, since they observed less than 10\% of female developers in the open-source projects analyzed. In response to this situation, Oliveira et al. \cite{oliveira2024navigating} 
have studied the path of women in software engineering, detecting persistent challenges in their career development. In the same line, Gila et al. \cite{genderEfects} described the effects of gender diversity on team performance based on personality types that show that female-gender developers looked uneasy in male-dominant teams. In contrast to this, and motivating the field of research,  recent findings  \cite{dowler2023diverse} show the clear benefits of gender-diversity in developing software products in other scientific fields \cite{yang2022gender}.

Also, regarding the race/ethnicity of the development team, Nadri et al. \cite{nadri2021insights} presents a novel qualitative analysis of non-merged pull requests and classifies the reasons why the pull requests were not merged based on four perceived racial identities: Asian/Pacific Islander, Black, Hispanic, and White. The authors found that perceptibly Black and Asian/Pacific developers had pull requests rejected at higher rates than those of perceptibly White developers. Other works as \cite{furtado2020successful} examine how geographic location associates with pull request submission and acceptance rates in open source projects. The authors found that most closed pull requests came from developers in the United States, the United Kingdom, and Germany, countries with a high HDI\footnote{The Human Development Index (HDI) is a measure of health, education, income and living conditions in a each country, compiled by the United Nations Development Programme.}. \change{Finally, several works, such as \cite{ortu2017diverse, shameer2023relationship}, mixed gender and nationality to investigate their relation within GitHub teams, analyzing how these factors vary across software projects and how demographic diversity can influence participation patterns and collaboration dynamics within software development teams.}  \looseness=-2

Disabilities have recently arisen as a focus of interest. For instance, recent research in the field has focused on creating more inclusive and diverse teams and integrating socially protected groups such as neurodivergent people. For example, Silva et al.~\cite{10.1145/3439961.3439986} identify the main techniques in Agile Methodologies to promote the inclusion of people with disabilities in development teams, and Liebel et al. \cite{liebel2024challenges} have proposed a set of recommendations to the industry to include developers suffering from ADHD (Attention-Deficit/Hyperactivity Disorder). Furthermore, answering the need for intersectional analysis of diversity, Prana et al. \cite{prana2021including} have analyzed the intersection of gender and region origin in the research community. This study considered the consistency of gender inequalities across different regions.

\subsection{Diversity comprises a wider definition of participants}
\label{sec:related:non-developers}

However, a wider range of profiles often participate in software projects. Recent research has highlighted the relevant role of these profiles in software communities~\cite{canovas2022analysis}: individuals commenting and reacting in open-source communities, pull request reviewers, bug reporters, documenters, translators, or developer advocates. In addition, recent studies about open-source communities \cite{miller2022did} have highlighted the toxicity in these communities, motivating the need to profile communities if we want to avoid toxic behaviors.  

In this sense, Paul et al. \cite{paul2019expressions} looked at code reviews, detecting that male reviewers were more likely to provide negative comments to female contributors. In contrast, Zolduoarrati and Licorish \cite{ZOLDUOARRATI2021106667} analyzed Stack Overflow archival data over 11 years, showing that female contributions tend to be more cooperative, supportive and collective and that women are more willing to share knowledge than their male counterparts. \change{Roles regarding the creation and maintenance of the documentation for software projects \cite{dagenais2010creating} and the translation of this documentation  \cite{arjona2013preliminary} have also been studied pointing out their relevance inside a software project.}  Another profile that has raised concerns in terms of diversity is that of modelers, and recent works \cite{liebel2024human} in the model-driven community also focused on the human aspects of modelers, emphasizing the modelers' diversity and highlighting the relevance of their background in modeling decisions.


\subsection{Diversity and human aspects impacting software end-users}
\label{sec:related:endusers}

Diversity and the human aspects of end-users play a critical role in software projects, as they directly impact usability, accessibility, and overall user satisfaction. \change{Seminal works on socio-technical systems have highlighted the relevant roles of the social context around software systems \cite{mumford2000socio}}, and have increasingly highlighted the importance of inclusiveness in software development, emphasizing that neglecting diverse user characteristics, limitations, and abilities can lead to dissatisfaction and exclusion. Studies such as those by \cite{grundy2021addressing} and~\cite{khalajzadeh2022diverse} propose taxonomies and analytical frameworks to better understand human-centric issues in software engineering. A key challenge is that inclusiveness is often overlooked during the development process, resulting in software that fails to accommodate the full range of human diversity \cite{khalajzadeh2022supporting, tizard2020voice}. For instance, by integrating user feedback throughout the life-cycle, developers can ensure that applications serve a broader audience while minimizing the risk of bias and accessibility barriers~\cite{hou2024evaluating}.

Efforts to adapt software to specific populations have already been made in the research community. The concept of Adaptive User Interfaces (AUIs) has gained traction in research \cite{miraz2021adaptive,10554790}, 
offering solutions that personalize interfaces for different user groups, including individuals with chronic conditions \cite{10554790}. Similarly, crowd-testing has emerged as a valuable approach in software development, enabling diverse users to participate in the testing phase and provide real-world insights into software usability. Studies on crowd testers, such as \cite{wang2022context,wang2019characterizing}, emphasize the importance of selecting a proper tester community to ensure a representative evaluation of software.

\subsection{Diversity in machine learning documentation frameworks}
\label{sec:related:ml}

Recent documentation and reporting frameworks in the Machine Learning (ML) field, such as Datasheets for Datasets \cite{gebru2021datasheets}, Data Statements \cite{bender2018data}, and Data Cards \cite{pushkarna2022data}, have become crucial to ensuring transparency, accountability, and fairness of these systems. These frameworks aim to capture the various aspects of data collection, model development, and evaluation while explicitly addressing biases, ethical considerations, and the societal impact of ML technologies. By incorporating diversity-related metadata into documentation, stakeholders can better assess whether ML systems fairly represent different demographic groups, mitigate potential biases, and adhere to inclusive algorithmic decision-making principles. Public regulatory initiatives such as the EU AI Act 
and the US AI Bill of Rights 
have started to adopt them, showing the increasing interest of public administrations in diversity.

A relevant contribution of these frameworks is their broader definition of the participants, which aligns with the perspective outlined in the previous subsections. They emphasize the role of individuals responsible for gathering and annotating training data, as well as those governing the project, highlighting the influence these groups can have on the final behavior of ML models. \change{However, in the broader context of software projects, their focus is limited, as they concentrate almost exclusively on datasets while overlooking other critical assets such as code repositories and project management plans. Moreover, the range of relevant participants differs: whereas annotators are central in ML projects, developers, testers, and end users often have a greater impact in software development. Thus, although these frameworks underscore the importance of diversity and have attracted attention from public institutions, they fall short of capturing the complexity of teams in software projects.}

\section{Current practices on reporting diversity in open-source software}
\label{empirical}
After analyzing the current state of the art on documenting diversity aspects in the previous section, we gathered insights into whether these aspects are present in Open-Source Software (OSS), a key field where they could and should be applied. 
OSS leverages the collaborative work of contributors to develop and evolve software projects. Transparency and openness are, therefore, key to keeping the project and the community alive, vibrant, and successful. Several works have emphasized the need to include the contributing guidelines \cite{elazhary2019not}, the code of conduct~\cite{tourani2017code} of these contributors, or explicitly define the governance of a project~\cite{izquierdo2023governance}, but no indications or techniques address diversity information. To evaluate the current practices, we conducted a study to analyze the presence of diversity information regarding the relevant groups identified in the previous section across OSS projects on \textsc{GitHub}.

\subsection{Study method}
\change{The method we followed to analyze current practices on reporting diversity in OSS included \changetwo{five phases, namely: repository collection, where we built a dataset of OSS projects for the study; analysis dimension definition, where we defined the dimensions to analyze practices on reporting diversity; prompt and tool refinement, where we designed an LLM-based approach to carry out the analysis; validation, where we performed the validation of the LLM-based approach; and repository analysis, where we conducted the analysis. In the following we describe each phase in detail, while next section reports the results of the analysis.}}

\medskip
\noindent\change{\textbf{Repository collection.}}
We studied the 200 top-starred repositories\footnote{In this section we use the terms project and repository as synonyms.} in \textsc{GitHub} for the top 5 programming languages\footnote{According to the ranking of top programming languages in \textsc{GitHub} in 2024.}, namely: C\#, Java, JavaScript, Python, and Typescript. We therefore analyzed a total of 1,000 repositories. To create the dataset of our study, we first retrieved the list of the top repositories for each selected programming language using the \textsc{GitHub} API on January 22$^\text{nd}$, 2025, and collected the first 200 repositories that  had recent activity in the form of commits within the last 30 days. Then, for each repository, we recovered all non-coding files aimed at documenting how the project was developed, specifically: readme, contributing, code of conduct, governance, code owners, community, support, security, release and funding.  For each file, we searched for different variations of their name, \emph{e.g.}, for codes of conduct we looked for files with names such as \texttt{coc.md}, \texttt{code\_of\_conduct.txt} or \texttt{CODE-OF-CONDUCT.adoc}. Finally, we analyzed the extracted files to search for diversity documentation.

\medskip
\noindent\change{\changetwo{\textbf{Analysis dimension definition.}}}
The search for diversity documentation evidence has focused on five main groups extracted from the review of the previous section, and further presented in Section \ref{sec:card}. Specifically:
(A1) the development team, and whether they mention profile aspects about the developers;
(A2) non-coding contributors, and whether they explain non-coding roles;
(A3) tests with potential users, and whether they mention labor force or report platforms;
(A4) usage context, including the target population and possible specific adaptations; 
and (A5) governance, including information on funders.
Areas A1 and A2 are motivated by the works on diversity for developers and non-coding roles (cf. Sections~\ref{sec:related:developers} and~\ref{sec:related:non-developers}), while A3 and A4 are inspired by the need for specifying human aspects (cf. Section~\ref{sec:related:endusers}).
The works on ML documentation (cf. Section~\ref{sec:related:ml}) prompted A5.
For each area, we reported whether the OSS project included (or not) such information.

\begin{lstlisting}[%style=promptStyle,
float=t!,
caption={Excerpt of the prompt used for searching evidence for A1 (\emph{i.e.}, Development Team).}, label=lst:prompt]
Analyze the following text and respond in JSON format:

1. Does the text mention the development team explicitly?
2. Are any specific aspects of the development team mentioned (e.g., team size, geographic diversity, gender diversity, roles, expertise, etc.)? If yes, list the aspects mentioned.

Output format:
{
  "mention_to_dev_team": "yes/no",
  "profile_aspects": {
    "mentioned": "yes/no",
    "aspects": ["list of aspects if mentioned, otherwise empty"]
  }
}
\end{lstlisting}

\medskip
\noindent\change{\textbf{Prompt and tool refinement.}}
To perform the analysis and search for diversity documentation, we built a tool based on Large Language Models (LLMs)\footnote{Tool's repository: \url{https://github.com/SOM-Research/Analyzer-Diversity-Card}}. In particular, we relied on GPT-4o-mini. 
The use of LLMs for analyzing text has demonstrated their effectiveness, for instance, for sentiment analysis~\cite{DBLP:conf/ucami/Carneros-PradoV23, Tran2024}, text classification~\cite{deng2023survey, sun2023text}, studying social interactions~\cite{DBLP:conf/daset/NadiNMANTASVR23}, and detecting ethical deviations~\cite{icseseis}.
In our context, the use of LLMs eliminates the need for extensive preprocessing, enabling a more streamlined
and direct analysis of the text.

\change{As a preliminary step before the automated analysis, we designed specific prompts for each of the five diversity areas considered (A1–A5).
The prompts were initially drafted by one of the authors and iteratively refined in collaboration with two others, using a sample of 20 representative projects.
This sample was used exclusively for prompt development and was not included in the final dataset.}
\change{Each prompt follows a few-shot strategy and includes five complete examples, composed of an input text fragment and its corresponding output in structured JSON format.
The examples cover a range of cases (i.e., from affirmative to negative and ambiguous) and illustrate different levels of detail regarding diversity-related aspects (e.g., team size, roles, gender, or geographic diversity).
This structure is intended to guide the model both in terms of the content to be identified and the expected format of the response\footnote{The full set of prompts is available at: \url{https://github.com/SOM-Research/Analyzer-Diversity-Card/blob/main/src/classifier/prompts.py}}}.
\change{Throughout the refinement process, both the wording and structure of the prompts were incresingly adjusted to promote consistency and reduce ambiguity in the model’s responses.
Specific measures were also introduced to mitigate potential hallucinations, including the rephrasing of unclear expressions and the enforcement of strict output formatting in JSON, ensuring reproducible and interpretable results.
A first evaluation on the 20 refinement projects continued until the model reached a performance level deemed satisfactory (above 90\%).} Listing \ref{lst:prompt} shows an excerpt of one of the five prompts used, in particular, the one addressing A1.

\medskip
\noindent\change{\textbf{Validation.}}
\change{Once the prompts and methodology were stabilized, a formal validation was conducted on a random sample of 50 projects from the full dataset.
For each project, the evidence extracted by the model across the five areas was manually reviewed and compared against the interpretation of the authors.
The validation was carried out by two authors, who independently reviewed the outputs and discussed discrepancies until reaching full agreement. This process yielded a 92.54\% agreement between the model’s output and human assessment, supporting the reliability of the tool for large-scale application.}

\medskip
\noindent\change{\textbf{Repository analysis.}}
\change{After the refinement and validation of the tool, we applied it to the full dataset of 1,000 repositories.
For each project, the extracted non-code files were provided as input, and the prompts guided the LLM-based tool to identify the presence (or absence) of diversity information across the five defined areas (A1–A5).
The classification results, together with the raw documentation data extracted from the projects, are openly available.\footnote{Zenodo DOI: \url{https://doi.org/10.5281/zenodo.14981689}}}



\begin{table}[t!]
\centering
\scriptsize
\renewcommand{\arraystretch}{1.3}
\begin{tabularx}{\textwidth}{X*{5}{R{3.6em}}R{3em}}
\multicolumn{1}{c}{\textsc{Area}} & 
\multicolumn{1}{r}{\textsc{C\#}} & 
\multicolumn{1}{r}{\textsc{Java}} & 
\multicolumn{1}{r}{\textsc{JS}} & 
\multicolumn{1}{r}{\textsc{Python}} & 
\multicolumn{1}{r}{\textsc{TS}} & 
\multicolumn{1}{r}{\textsc{Avg}} \\
\toprule
A1 -- Refer to development team                        & 28.0\% & 31.5\% & 28.0\% & 27.5\% & 33.0\% & 29.6\% \\
\hspace{2.1em} \emph{...and mention profile aspects}   & 22.0\% & 20.5\% & 22.0\% & 21.5\% & 26.0\% & 22.4\% \\
\midrule
A2 -- Describe non-coding contributors                 & 30.5\% & 27.0\% & 32.0\% & 27.5\% & 31.5\% & 29.7\% \\
\hspace{2.1em} \emph{...and explain non-coding roles}  & 24.0\% & 20.5\% & 26.0\% & 18.5\% & 24.4\% & 22.7\% \\
\midrule
A3 -- Tests with Users                                 &  3.5\% &  3.5\% &  1.5\% &  5.0\% &  4.0\% &  3.5\% \\
\hspace{2.1em} \emph{...and mention labor force}       &  0.0\% &  0.5\% &  0.0\% &  0.5\% &  0.5\% &  0.3\% \\
\hspace{2.1em} \emph{...and report platforms}          &  0.2\% &  0.2\% &  0.0\% &  0.3\% &  0.1\% &  0.2\% \\
\midrule
A4 -- Define specific use case                         & 84.0\% & 74.5\% & 78.5\% & 83.0\% & 68.0\% & 77.6\% \\
\hspace{2.45em}Specify target population               & 34.0\% & 28.5\% & 32.0\% & 36.5\% & 30.0\% & 32.2\% \\
\hspace{2.45em}Describe specific adaptation            & 15.0\% & 13.5\% & 17.0\% & 21.5\% & 15.5\% & 16.5\% \\
\midrule
A5 -- Define governance participants                   & 28.5\% & 31.5\% & 27.0\% & 27.0\% & 30.0\% & 28.8\% \\
\hspace{2.4em}Indicate funders                         & 27.0\% & 15.0\% & 32.0\% & 19.5\% & 30.0\% & 24.7\% \\
\bottomrule
\end{tabularx}
\caption{Presence of diversity documentation in 1,000 OSS projects.}
\label{tab:analysis}
\end{table}

\subsection{Study results}

Table~\ref{tab:analysis} shows the results of our study. 
The first column shows the area under analysis, while the rest of the columns indicate the percentage of projects including evidence for such areas. 
Columns 2 to 6 show the percentage for the subset of projects of a specific programming language, while the last column shows the average value.  
Rows show the results per area. 
Note that A1, A2 and A3 include subareas, for instance, A1 seeks evidence referring to the development team (\emph{e.g.}, 28.0\% of the C\# projects), and then, for those projects including information about developers, it searches for mentions to profile aspects 
(\emph{e.g.}, 22.0\% of the C\# projects referring to the development team).
As can be seen, the highest values are found for A4 (\emph{i.e.}, definition of specific use cases) but, in general, the presence of diversity information for all areas is low. 
In the following, we analyze the results per area. 

Information about the development team and non-coding contributors has been found in less than a third of the analyzed projects (\emph{i.e.}, 29.6\% and 29.7\% of the projects, respectively). 
However, when this information is available, it generally also includes additional details on development profiles or non-coding roles  
(\emph{i.e.}, 22.4\% and 22.7\% of the projects reporting information on the development team and non-coding contributors, respectively). 
Further analysis on what development profiles are reported revealed that the main information facilitated is related to development roles and team size, while geographic and gender diversity are rarely commented. 
Regarding non-coding roles, the three most used roles are reporters, translators, and \emph{documenters}. 
Information about governance and funders shows results similar to those obtained for the development team and non-coding contributors, with less than a third of the analyzed projects covering this area.



Evidence of references on user testing activities are the lowest ones, averaging 3.5\% across projects. 
When this information is reported, details on the use of labor force and reporting platforms are nearly absent 
(\emph{i.e.}, 0.3\% and 0.2\% of the projects reporting information on testing, respectively). 
The presence of information related to usage context (A4) is the highest compared to other areas, with specific use cases being the most documented, averaging 77.6\% across projects. 
A detailed analysis revealed that the most common terms associated with use cases include concepts such as software creation, data handling, application development, tool building, and web solutions. 
These results highlight a strong focus on general-purpose applications.
Regarding target populations, the information is limited and present in only 32.2\% of projects. 
Among these, developers are by far the most frequently mentioned group, followed to a lesser extent by other technical profiles (\emph{i.e.}, engineers, scientists). 
Data on specific adaptations is the most scarce, and is mentioned in only 16.5\% of projects, where most references are related to language localization.

\changetwo{We believe these findings reveal a general lack of documentation on diversity, confirming our initial hypothesis and underscoring the need for mechanisms that support its explicit definition. The results also align with trends in the scientific literature, such as a strong focus on generic use cases, limited descriptions of target populations and adaptations, and scarce visibility of funding sources. These insights motivate the creation of a repository-level Software Diversity Card to make such information explicit and comparable across projects.}

\changetwo{In OSS repositories, documentation often prioritizes what the software does and how to use it, while intended audiences appear less frequently \cite{Prana2019Readme}. Contribution guidelines and codes of conduct are common, yet governance and funding disclosures remain inconsistent and are sometimes externalized \cite{elazhary2019not,tourani2017code,izquierdo2023governance}. Overall, the findings indicate a need to better structure this information, emphasizing not only developer teams but also the governance and human aspects of software projects. In light of these results, the next section presents a conceptual proposal outlining the key dimensions of the Software Diversity Card.}


\subsection{\change{Threats to validity of the study}}

We also acknowledge that our results are subjected to a number of threats to validity.  Regarding the internal validity, which refers to the inferences we have made, we relied on the data provided by the GitHub API. 
These data may not be complete and/or include toy projects (\emph{e.g.}, projects addressing homework assignments). \change{It is also important to note that even if the project indicates that it uses a specific programming language, it may not qualify to be considered as a software project (e.g., a project with a list of resources that also includes Python scripts).}
\change{Other additional dimensions, such as geographic and linguistic differences among the projects may also become a threat to the internal validity.}
\change{To minimize this threat, we analyzed the top repositories of each programming language, which should include reference projects in the platform.}
Regarding the external validity, our data was collected on January 22$^\text{nd}$, 2025; and therefore our results should not be directly generalized to any project without proper comparison and validation.\looseness=-1

\section{Dimensions of the diversity card}
\label{sec:card}

\begin{figure}[b!] \centering
\includegraphics[width=0.7\columnwidth]{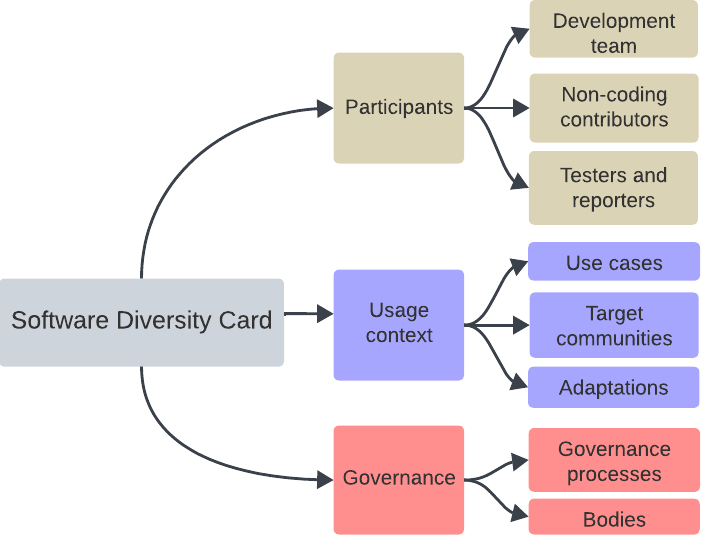}
 \caption{Overview of the Software Diversity Card.}
 \label{fig:feature}
\end{figure}

\change{This section introduces the structure of the Software Diversity Card, which we organized into three main parts; Participants, Usage Context, and Governance. In Figure \ref{fig:feature} we can see an overview of the card structure.}

\change{The proposed separation is proposed as a way of distinguishing between endogenous and exogenous determinants of software projects within a socio-technical \mbox{system \cite{mumford2000socio}}. Participants correspond to endogenous agents whose activities, decisions, and interactions are generated within the boundaries of the project and directly affect its evolution, while the Usage Context (end users and broader social environment) corresponds to exogenous factors that constrain or shape project outcomes. Governance serves as a mediating layer, operating as a boundary-setting mechanism that connects internal agency with external conditions \cite{kallinikos2004social}. This tripartite distinction aligns with long-standing perspectives in socio-technical systems research \cite{baxter2011socio, geels2004sectoral, mumford2000socio}), which emphasize that successful systems design requires balancing the technical core (the participants’ actions), the social context (usage context), and governance.  }

\subsection{Participants in software projects}

This part aims to profile the different types of participants behind developing and maintaining a software project. For each of these teams, and inspired by recent works in software diversity and inclusion \cite{bjorn2023diversity, albusays2021diversity}, the goal is to document a minimum set of demographic attributes such as age, perceived gender, ethnicity, location, and disabilities; and cultural characteristics of the team, such as the spoken language, the socio-economic status, the job title, the experience, and education level, among others. The most clearly identifiable team behind a system is the \textit{development team}. Several studies have studied how diverse teams are beneficial for the outcome of software systems, and this information would be valuable for end-users to trust in the software (\emph{e.g.}, the case of the period tracker app), or for public regulators to ensure the app is developed in a certain region 

However, there are other types of teams involved in the development of software products that also play a pivotal role.  \change{\textit{Non-coding contributors} have largely been noticed as relevant in the evolution and maintenance of open-source software projects~\cite{canovas2022analysis, santos2013attraction}}. For example, roles like the \textit{issue reporters} are influential in the evolution of software projects. However, social groups less familiar with reporting issues may see their needs under-represented during the evolution of software systems \cite{tizard2020voice}. To this end, gathering diverse information about reporters could help companies avoid this kind of bias.  

On the other hand, \textit{translators, issue reporters} and \emph{documenters} 
are influential actors in creating the community around software projects and, in most cases, are in charge of reaching new layers of potential users. For instance, open-source communities have been calling for translating contributions in order to make their software more accessible\footnote{Drupal's community call for documentation translators: \url{https://www.drupal.org/community/contributor-guide/contribution-areas/translations}} \footnote{Open Stack's community call for documentation translators: \url{https://docs.openstack.org/contributors/common/i18n.html}}. Besides, public \textit{reporting users} or user feedback can also play a relevant role in evaluating a software product \cite{tizard2020voice}. For example, in the video game industry, public reporting users are a valuable asset for analyzing the success of a video game \cite{yu2023mining}. Despite this feedback being often considered the \mbox{``voice of the users'' \cite{tizard2020voice}}, only a fraction of the software's users indeed provide it. If the demographics of these users were not representative of the software target users, there may be a generated biased software evolution against these under-represented groups, similar to the case of issue reporters.

On the other hand, public and controlled \textit{beta testing}, along with \textit{crowd testing experiments}, are frequently employed in software development to gather pertinent feedback from potential users \cite{leicht2017leveraging}. While in controlled public beta testing the users can be self-selected, in crowd-testing experiments groups of individuals are hired to test the software. The feedback gathered for these experiments will help product owners evaluate the software's suitability. Still, it can also be helpful for end-users and public administration when assessing the adoption of new software. However, the quality of this feedback depends on many factors, such as the diversity of testers and specific instructions for the tasks to be performed. Other factors include the time dedicated to test the software, the labour conditions, the app's maturity level, and the number of iterations. These considerations have been identified in works in the machine learning field (such as Crowdworksheets \cite{diaz2022crowdworksheets}) and also in the \mbox{Software Engineering field \cite{leicht2017leveraging,TSAI2023107103}}.

\subsection{The usage context of the software}

To properly understand why a specific composition of actors is relevant, we need information about the \textit{social context} where a system is intended to be used. Our proposal allows reporting the particular use cases the software is designed for and the social and cultural constraints that are applicable to these use cases. Following the example of the period tracker app, this app could be designed for young women living in Catalonia, a region of Spain with a specific co-official language (Catalan). This social context may generate the need to adapt the app's user interface to Catalan and test the app on Catalan speakers to ensure it fits their needs.

However, the usage context may vary depending on the app's potential target users. For instance, a public administration or a product owner would require the app to be accessible for a particular social group (\emph{e.g.}, people with disabilities, elderly people, minority languages, neurodiverse people). In that sense, our proposal allows defining specific target populations (such as deaf people) and the particular adaptation of the app to reach those users (\emph{e.g.}, videos with audio descriptions). 
Each adaptation could modify or require specific testing conditions, such as ensuring accessibility in the tester's device.

\subsection{The governance of the software project}

High-level information about the governance process and funders of the software products is also relevant for users to better understand the software such as the process followed to prioritize features or bug changes, or the criteria used to accept contributions. Moreover, this information is often hidden from the final users, and the recommendation to disclose it could help in creating a culture of better software practices. To this end, the goal of this part of the diversity card is to document information about the type of governance processes the project has (\emph{e.g.}, private, corporation, public, open community, research project), who are the main decision-makers and how decisions are made, the legal regulation the product is under, information about the funders of the project (\emph{e.g.}, private, a public administration, or an NGO), and information about the shareholders.

\section{Modeling diversity in software projects}
\label{sec:model}


Our diversity card proposal aims to serve as a communication asset for software owners, end users, and public regulators. To promote its adoption and ensure consistent use, we need to provide structured resources and integrate them with existing process management tools in software development initiatives. In this section, we introduce a domain-specific language (DSL) designed for documenting software diversity, along with a toolkit to assist in the creation of Software Diversity Cards for various user profiles. We believe that this formalization offers a common language that enables stakeholders involved in a software project to understand, communicate, and discuss diversity-related aspects effectively.

In this section, we first present the abstract syntax of the DSL, which formalizes the concepts outlined in Section~\ref{sec:card}\changetwo{, following the UML standard. Although many attributes are optional for most projects, they are not indicated in the metamodels to improve their readability. Note that the \texttt{*} cardinality in an association end indicates that there may be no informed associated instances.}

Then, we introduce two different concrete syntaxes, one tailored for users with coding skills, implemented as a language plugin for Visual Studio Code, and a second one oriented to non-coders based on a web-based tool that allows non-technical people to create cards easily. \changetwo{These concrete syntaxes provide users with full flexibility in reporting the information they have available and willing or enforced to share.}

\subsection{Abstract syntax}

The metamodel is organized into packages that mimic the required dimensions of a Software Diversity Card (see Section~\ref{sec:card}). In the following, we present the elements of each package.

\subsubsection{Participants}

\begin{figure}
\centering
\includegraphics[width=0.9\textwidth]{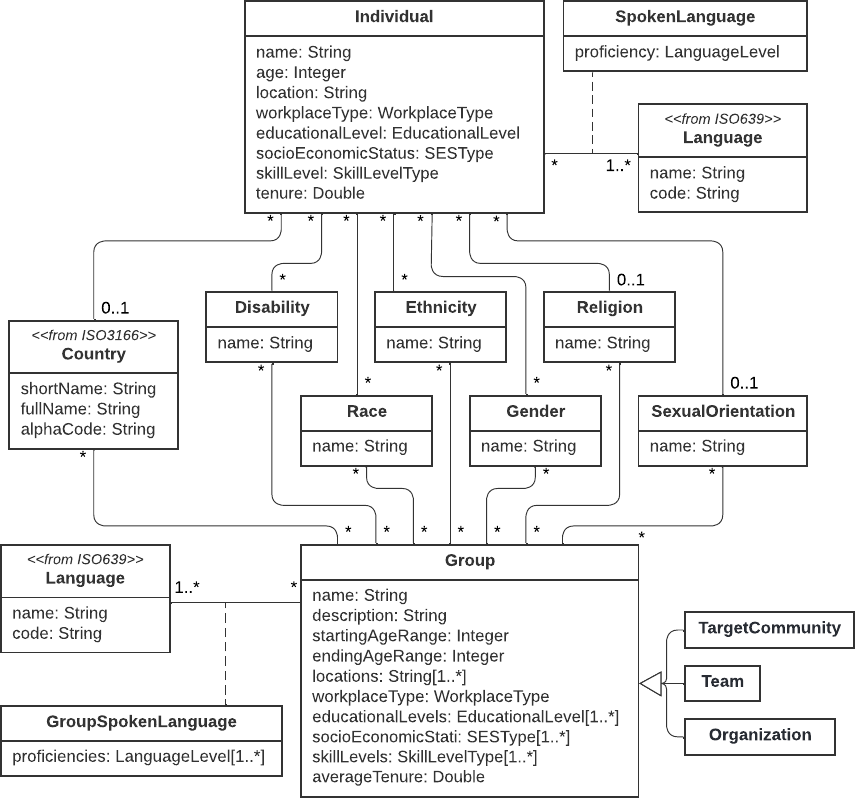}
 \caption{Metamodel for collecting data about individuals and groups.}
 \label{fig:mm_entities_individuals}
\end{figure}

In Figure~\ref{fig:mm_entities_individuals}, we show an excerpt of the metamodel that collects data about individuals and the different groups engaged in a software development project. The main attributes collected in this model are inspired by recent works in software diversity and inclusion~\cite{bjorn2023diversity, albusays2021diversity}, as mentioned in previous sections.

\texttt{Individuals} \change{are a generic representation of the common data to be collected for} both team participants and other individual stakeholders in software projects. \texttt{Individuals} report their \texttt{spokenLanguages} (including their respective \texttt{proficiency} levels, based on the Common European Framework of Reference for Languages\footnote{\url{https://www.coe.int/en/web/common-european-framework-reference-languages}}), \texttt{ethnicity}, \texttt{gender}, \texttt{age}, \texttt{socioEconomicStatus} (like the Kuppuswamy socioeconomic status class classification~\cite{lakhumna2025socio} and similar), \texttt{skillLevel} and \texttt{tenure}, among other information. On the other hand, \texttt{Groups} include homologous data, but in aggregated format, \change{in order to uniquely specify the generic data that is collected for representing groups of individuals. For instance, a project may report an aggregated profile of the development team instead of the profile of each of their individual members.} In that sense, \texttt{Groups} represent the essential piece of our model, and \texttt{Teams}, \texttt{Organizations}, and \texttt{TargetCommunities} are the specific types of \texttt{Groups} that compose the card.

\begin{figure*}
\centering
\includegraphics[width=0.82\textwidth]{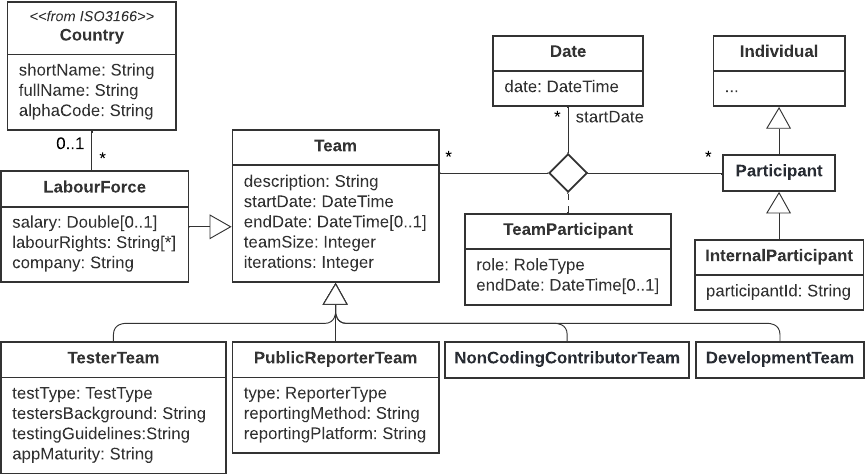}
 \caption{Participants and teams hierarchies metamodel.}
 \label{fig:mm_teams_participants}
\end{figure*}

In Figure~\ref{fig:mm_teams_participants}, we can see an excerpt of the team hierarchies and its participants. \change{\texttt{Teams} are classified into two orthogonal specialization sets. The first set specializes a \texttt{Team} according to its objectives and responsibilities, capturing its functional role within the system. \texttt{Teams} are therefore classified as either \texttt{TesterTeams}, \texttt{PublicReporterTeams}, \texttt{NonCodingContributorTeams}, or \texttt{DevelopmentTeams}; each of them contributing with specific information. The second specialization set classifies \texttt{Teams} based on organizational sourcing, distinguishing between internal and external \texttt{LabourForces}, for which further details regarding contractual aspects such as \texttt{salary} and \texttt{labourRights} (primarily according to the \texttt{Country} where the team is based in), and the \texttt{company} the team belongs to is informed. These two specialization sets are independent, in order to comprehend diverse combinatorial scenarios. For instance, a \texttt{TesterTeam} could be outsourced as a \texttt{LabourForce}, whereas a \texttt{DevelopmentTeam} could be internal for the development of a software system.}

\change{In addition, \texttt{Participants} are a specialization of \texttt{Individuals}, so as to distinguish the concrete actors that are associated with \texttt{Teams}.} 
\texttt{Participants} belong to a \texttt{Team} within a concrete date range, and exercise a particular \texttt{role} (\emph{e.g.}, reporter, developer or reviewer, among others) in such period. Those \texttt{Participants} who are affiliated to the organization responsible for the development may inform their \texttt{participantId} (\emph{i.e.}, their internal business identifier; \emph{e.g.}, to enable them to access the organization's owned systems) \change{and are represented as \texttt{InternalParticipants}.}

\subsubsection{Usage context}

\begin{figure*}[h]
\centering
\includegraphics[width=0.8\textwidth]{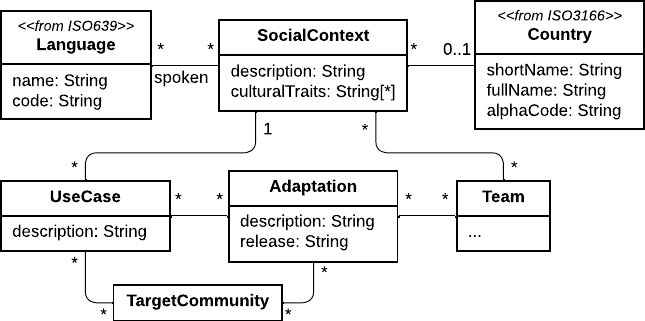}
 \caption{Usage context metamodel.}
 \label{fig:mm_context}
\end{figure*}

As mentioned in previous sections, software systems must consider its usage context, captured in Figure~\ref{fig:mm_context}. In that sense, the \texttt{SocialContext} reflects the environment of social interactions, \texttt{spokenLanguages}, and \texttt{culturalTraits} where a software system is going to be exploited, in order to effectively address the users' needs and nuances. The features of a system are built to respond to the different intended \texttt{UseCases} and their respective \texttt{TargetCommunities} or end users. For particular \texttt{TargetCommunities} who have specific needs or alternative, complementary requirements, it may be necessary to make additional \texttt{Adaptations} delivered in a \texttt{release}.

\subsubsection{Governance}

\begin{figure*}[h]
\centering
\includegraphics[width=0.9\textwidth]{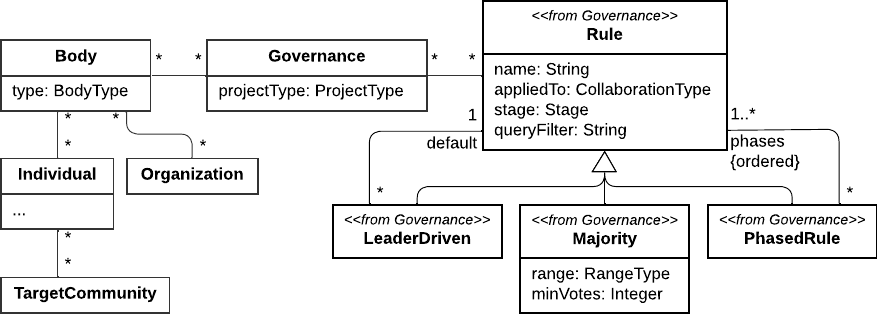}
 \caption{Governance metamodel.}
 \label{fig:mm_governance}
\end{figure*}

\texttt{Governance} is a key facet that influences diversity and inclusion in a project, modelled as in Figure~\ref{fig:mm_governance}. The Software Diversity Card may contain information in regard to the \texttt{Bodies} (\emph{e.g.}, the board of administrators) that govern the system construction, which are composed of either individuals and/or organizations. Governance \texttt{Rules} in software development projects establish clear guidelines for collaboration between core and external contributors, ensuring they work effectively together to advance the project throughout its entire lifespan. \texttt{Rules} can be designed using the DSL from \cite{CanovasGovernanceRules}.

\subsection{Tool support}

In this section, we lay out the details of a supporting toolkit that implements a concrete syntax of the DSL and provides transformations to JSON and Markdown formats. This allows any user to implement form-based or diagramming solutions to facilitate the design of software diversity cards, integrate such information within identity and management platforms, and document diversity aspects of a system in online platforms. 

\subsubsection{Textual concrete syntax}

We created a Visual Studio Code extension\footnote{Extension home repository: \url{https://github.com/SOM-Research/SoftwareDiversityCard}} that offers users an editor for completing software diversity cards. The concrete syntax is implemented using Langium\footnote{Langium's homepage: \url{https://langium.org}}, an open-source tool for defining DSL grammars and enforcing well-formedness rules. The extension includes features such as syntax highlighting, syntax checking to ensure proper DSL instantiation, validation rules for model consistency, and autocomplete suggestions to simplify the association of entity instances. To facilitate its accessibility, the extension has been published in the Visual Studio Code Marketplace\footnote{VSCode Marketplace page of the extension: \url{https://marketplace.visualstudio.com/items?itemName=SOMResearchGroup.SoftwareDiversityCard}}.

\begin{lstlisting}[language=grammar, caption=Excerpt of the grammar implemented in Langium., label=list:grammar]
grammar SoftwareDiversityCard

entry Model: [...]

SocialContext: 'socialContext' name=ID
    'description:' description=STRING
    ('country:' country=ISO3166)?
    ('spokenLanguages:' '[' (spokenLanguages+=ISO639)
        (',' spokenLanguages+=ISO639)*']')?
    ('useCases:' '[' (useCases+=[UseCase:ID])
        (',' useCases+=[UseCase:ID])*']')?
    TeamsArray;

Adaptation: 'adaptation' name=ID
    'description:' description=STRING
    ('release:' release=STRING)?
    ('targetCommunities:'
        '['(targetCommunities+=[TargetCommunity:ID])
        (',' targetCommunities+=[TargetCommunity:ID])*']')?
    TeamsArray;

Participant: 'participant' name=ID
    'age:' age=NUMBER
    'location:' location=STRING,
    [...]
    ('participantId:' participantId=STRING)?;

Team: (TesterTeam | DevelopmentTeam |
      NonCodingContributorTeam | PublicReporterTeam);

fragment TeamAttributes: name=ID Entity
    'description:' description=STRING
    'startDate:' startDate=DATE
    ('teamSize:' teamSize=NUMBER)?
    [...]
\end{lstlisting}

A snippet of the grammar is shown in Listing~\ref{list:grammar}. For usability and maintainability reasons, we have designed all relationships as uni-directional. We have defined Langium ``terminals'' for all enumerations included in the DSL, as well as for the ISO codes of countries and languages (\emph{i.e.}, \texttt{ISO3166} and \texttt{ISO639}, respectively), to ease the introduction of such data. Inherited attributes from entities and teams are established as Langium ``fragments'' (\emph{e.g.}, \texttt{TeamAttributes}), which allow a single definition of shared elements.

\begin{figure*}[b!]
\centering
\includegraphics[width=0.85\columnwidth]{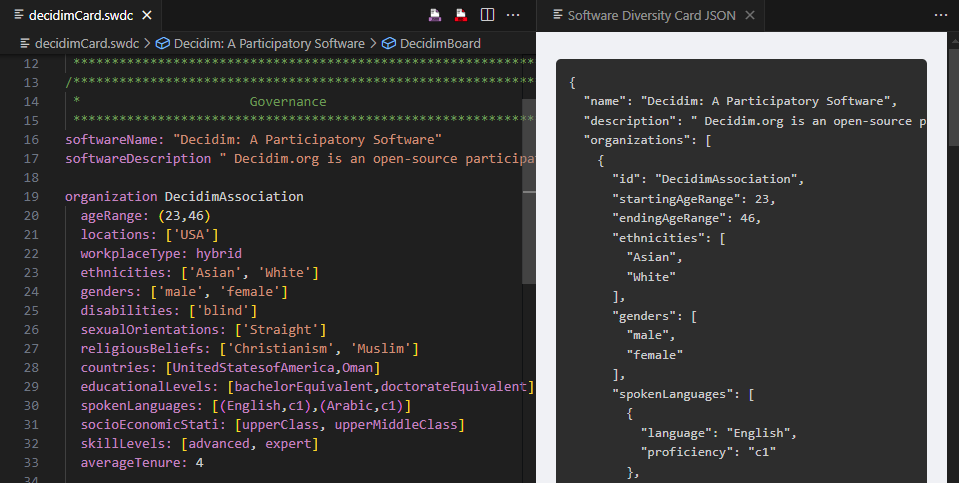}
 \caption{VSCode extension UI with a Software Diversity Card and the generated JSON output.}
 \label{fig:toolsupport_vscode_plugin}
\end{figure*}

An excerpt of a Software Diversity Card compiled using the extension is displayed in Figure~\ref{fig:toolsupport_vscode_plugin} (left). Pair values are delimited by ``('' and ``)'' (\emph{e.g.}, \texttt{(English,c1)} in \texttt{organization.spokenLanguages}); arrays are delimited by ``['' and ``]'' (\emph{e.g.}, \texttt{organization.ethnicities}); and linked elements are referenced by their \texttt{id}.

The Visual Studio Code extension provides capabilities for exporting software diversity cards to JSON and Markdown (MD) formats. The information contained in the JSON file could be processed to populate extended properties in directory services (\emph{e.g.}, Active Directory, OpenLDAP and Google Cloud Identity) or identity and access management tools (\emph{e.g.}, Azure AD and platforms compatible with SAML or OpenID Connect). On the other hand, the MD file could be included as part of the initiative documentation in online repositories such as \textsc{GitHub}. 

An example of JSON code is depicted in Figure~\ref{fig:toolsupport_vscode_plugin} (right), following a typical block-based textual representation. Our plugin helps to simplify the creation of such a JSON description, and prevents manual errors such as introducing grammatical inconsistencies or missing required information.

\subsubsection{A form-based web tool for creating software diversity cards}
 \begin{figure}[b!] 
\centering
\includegraphics[width=0.8\columnwidth]{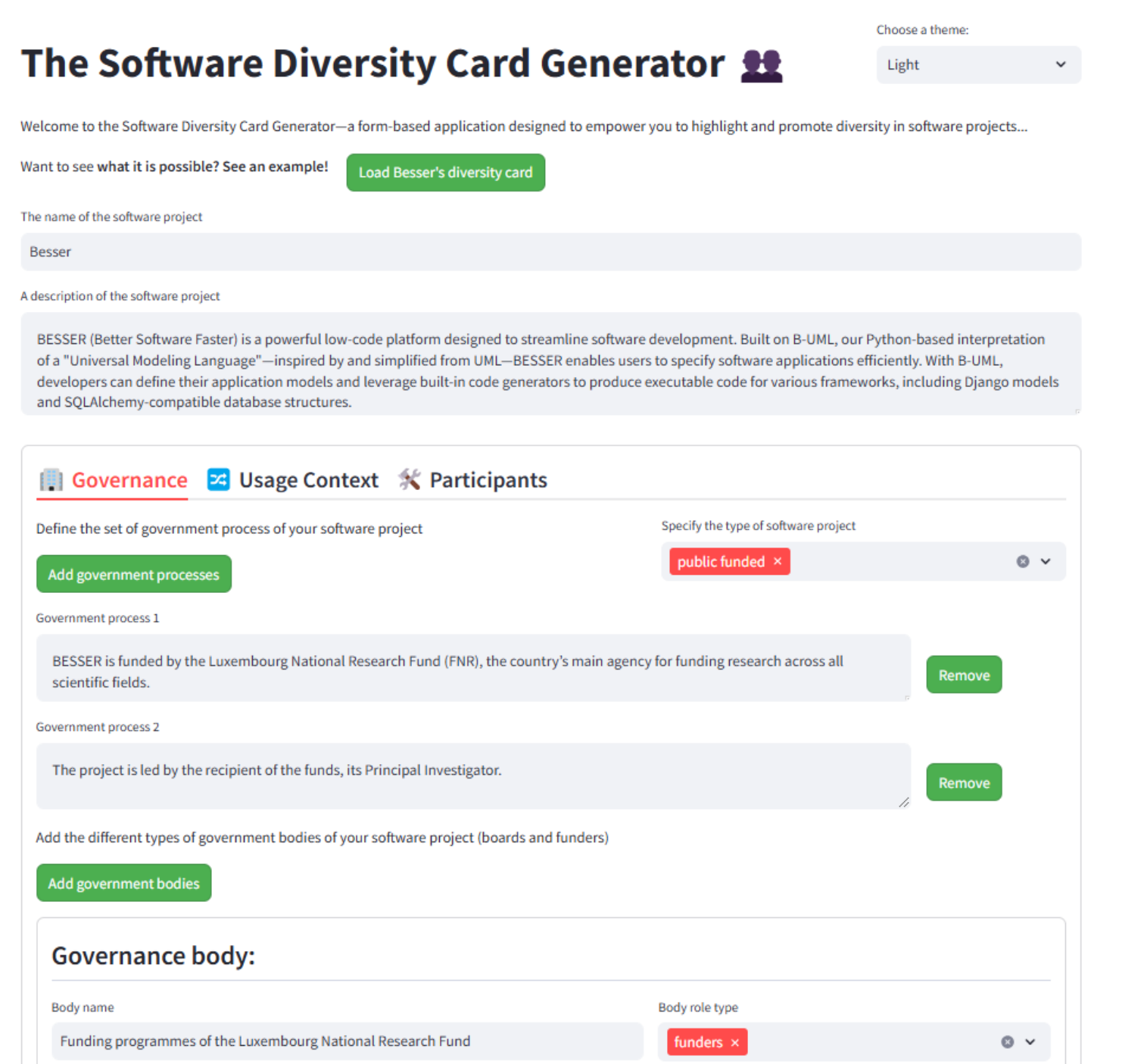}
 \caption{UI of the form-based web tool for creating software diversity cards.}
 \label{fig:builder}
\end{figure}

This section presents a web-based form tool to assist users in creating the cards, leveraging the previously introduced abstract syntax. Built with Streamlit\footnote{Streamlit project homepage: \url{https://streamlit.io}}, the tool is open-source and published alongside the language plugin for Visual Studio. The tool is web-form-based, meaning that the concepts of the abstract syntax are translated into forms that the user should fill in to complete the tabs, and is focused on aiding non-technical users in card's creation. Figure \ref{fig:builder} shows an excerpt of the UI of the tool and an online demo has been published\footnote{Online demo of the web editor: \url{https://huggingface.co/spaces/JoanGiner/SoftwareDiversityCard-Generator}}. 

One of the main advantages of this form-based tool is that it guides the user about each of the fields available and their expected format. For instance, as we can see in the figure, in the governance part, the user can create as many bodies as they need, assign the type of body using an option selector, and define, if needed, basic demographic information for each body. Another advantage of this web-based tool, compared to the presented language plugin,  is the smooth installation step, as this can be accessed directly from the web browser, which is ideally for non-coding users.

However, although this tool should be the preferred one to start creating cards, it has some limitations that make using the language plugin more suitable for specific use cases. For instance, it is challenging to manage a large number of participants as the form could quickly get overwhelming for the users. The same happens with attributes with a lot of nested dimensions (such as defining the level of knowledge of a language of participants of the development team) that become difficult to manage using a web-based form. Finally, once the users have filled out the web form in the app, the tool generates an equivalent Markdown and JSON representation of the cards. So the users can download and share the generated files with the community or the relevant stakeholders of the project.

\change{\subsection{DSL evaluation} }

\change{To evaluate the usefulness, completeness, and usability of the proposed DSL and its supporting tools, we conducted an empirical validation with experienced software developers. The experiment comprised a reading and a writing exercise. Initially, participants used the VSCode language plugin to read a complete diversity card and answered questions to measure the readability of the DSL. Subsequently, using the form-based web tool, they completed a partially filled-out diversity card, accompanied by a document containing project diversity information in natural language. This part assessed the usability of the tool and the identification of complex concepts, such as target communities and user reporting platforms. Post-exercises, participants took part in a semi-structured interview, providing quantitative feedback via Likert-scale items across criteria such as (1) usefulness (ability to capture diversity information), (2) completeness (adequacy of the DSL constructs), and (3) usability (ease of use and adoption potential); and qualitative insights through open questions within each category.}

\change{The experiments took place in August and September 2025 in Spain and Luxembourg, with 13 English-fluent participants composed mainly of research software developers from three research groups, with a median of four years of experience. The complete questionnaire and detailed study results are available in an online repository\footnote{The documentation can be found in the \textit{docs} folder of the tools repository: \url{https://github.com/SOM-Research/SoftwareDiversityCard}}. }

\change{The reading exercise results show that 93.5\% of responses were correct, indicating a high level of DSL readability. Most participants successfully identified key concepts such as target communities and reporting platforms. Incorrect answers were primarily provided by non-developers, which suggests that limited experience with VSCode may have introduced a bias. Divergent interpretations emerged around governance: while some participants recognized funders (e.g., the Barcelona Council) as relevant actors, others referred only to the project's board.}

  \begin{figure}[b!] 
\centering
\includegraphics[width=0.8\columnwidth]{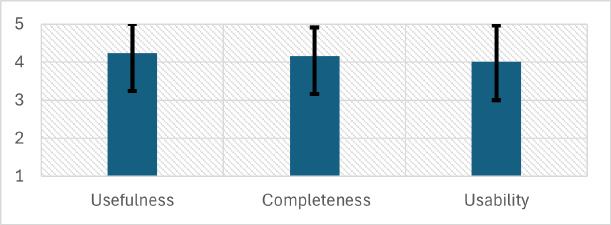}
 \caption{Results of the questionnaire grouped by the assessment criteria. Scores range from 1 (strongly disagree) to 5 (strongly agree). Usefulness comprises Q1, Q2, and Q3; Completeness comprises Q4, Q5, Q6; and Usability comprises Q7, Q8, and Q9 of the questionnaire.}
 \label{fig:questions}
\end{figure}

\change{The results of the writing exercise and post-exercise interviews are summarized in Figure~\ref{fig:questions}. The Likert-scale questions, grouped by evaluation criteria, reflect a positive assessment of the DSL and its supporting tools, with mean scores above 4 (on a 1–5 scale) for usefulness, completeness, and usability. In contrast, responses to the open-ended questions yielded several areas for improvement. Participants frequently mentioned compatibility issues with browsers and mobile devices, underscoring the need to adapt the tool for diverse platforms. They also recommended adding intersectionality options (e.g., combining gender and race) to balance privacy with informativeness. Additional suggestions included enriching the tool with graphical elements, such as maps, to support visualization of geographic distribution, and linking user profiles to specific roles in software projects (e.g., managers or developers). The latter could be realized through integration with business process management tools, as discussed in Section~\ref{sec:conclusions}.}

\change{\section{Use cases}\label{sec:usecases}}

In this section, we describe two mature open-source projects involving different groups of people using the proposed software diversity card. The two projects are Decidim \cite{aragon2017deliberative}, a participatory software used by over 400 administrations worldwide to allow citizens to participate in public decisions, and Besser \cite{alfonso2024building}, a low-code platform designed to enable non-coding profiles to develop software applications in various domains where one of the authors of this work is directly involved. Despite their distinct characteristics, such as target users, both projects are open-source, publicly funded, and part of a user community.

As a first step in describing the projects, we reviewed the online documentation of each project and attempted to complete the cards as much as possible. Then, we interviewed the project's development and product teams, presenting our proposal and asking about the card's dimensions. This section summarizes the key points of each project's card and discusses the challenges we and the software projects encountered in gathering the required card dimensions. The complete examples of the cards are available online\footnote{The card's examples are in the \textit{/examples} folder of the tools repository \url{https://github.com/SOM-Research/SoftwareDiversityCard}}, \change{and Tables \ref{tab:decidim} and \ref{tab:bessercom} summarize the overall completeness of the card for each use case.}

\subsection{Decidim: an open-source participatory democracy platform}

Decidim is an open-source platform that fosters participatory democracy and involves different groups \emph{participants}. The platform's \emph{development team} comprises maintainers, developers, and external contributor teams focused on specific features. Despite the extensive project documentation and clear commitment to transparency, we have found that the developers' profiles were not publicly disclosed. During the interview with Decidim, we learned that Decidim is composed of internal and external developer teams. In that sense, they pointed out that the external teams change over time, making it difficult for them to publish a long-standing team's profile information. \change{This evolving nature of the development team represents a challenge for the proposed card, in that specific case, we decided to capture the current state of the team, which can be seen in the complete example published online.}

Decidim has a community of \emph{non-coding contributors} who report bugs, propose features, and translate the software via its own online social platform, Metadecidim\footnote{Metadecidim home page: \url{https://meta.decidim.org}}. However, following the approach of many open-source communities, they adopted the ``anonymity first'' principle, where any user contribution only needs to be associated with a username. \change{While this approach aims to avoid bias and promote equal participation, it is challenging for them to profile the community of non-coder contributors accurately, and it represents an open challenge for the Software Diversity Card.} In that sense, we have been able to profile the size, number of contributions, and its expertise in the reporting platforms of both teams, that can be seen in the complete example published online. 

Regarding \textit{testers and public reporters}, they did several usability tests and public interviews with citizens in several participatory processes they drive. However, none of these studies have been made public or reported in the extensive project's documentation. They noted that this is a path of improvement for the project, as disclosing this information may help, for instance, in the adoption of Decidim by other institutions. However, they pointed to some limitations regarding this issue, such as the lack of UI/UX experts to build systematic tests and standard guidelines to disclose this information in open-source projects. \change{ This example illustrates one of the benefits of standardization efforts in this area, where the lack of guidance hinders the disclosure of diversity-related information.} 

\begin{table}[b!]
    \centering
\scriptsize
\renewcommand{\arraystretch}{1.3}
    \caption{Overview of the completeness of Decidim's card.}
    \label{tab:decidim}
    \begin{tabular}{|ll|l|}

       \hline
        \textbf{Dimensions} & \textbf{Team type} & \textbf{Profile completeness} \\ \hline   
        {\textbf{Participants}} & \textit{Developer Team} &  High as a summary of internal and external teams \\ 
        \textbf{} & \textit{Non-coding contributors}  & Limited due to OSS community (anonymity-first)  \\ 
        \textbf{} & \textit{Testers}  & High as extracted from public interviews  \\ 
        \textbf{} & \textit{Reporters}  & Limited due to the anonymity-first approach \\ 
         \hline
        {\textbf{Usage Context}} & \textit{Target Communities}  & Limited to the project's use cases \\ 
  
        \hline
      
        {\textbf{Governance}} & \textit{Boards}  &  High as is public information \\ 
       \textbf{} & \textit{Bodies}  &  High as they are public institutions  \\ %

        \hline

    \end{tabular}

\end{table}

\begin{lstlisting}[
label=lst:decidim,
language=DSL,
float=t,
upquote=true,
caption=Excerpt of the usage context part of the Decidim's card.,
   ]
targetCommunity nonDigitalSkilled
  description: "Elder citizen or citizen with low digital skills..."
  ageRange: (60, 100)
  locations: ["Barcelona"]
  workplaceType: presential
  countries: [Spain]
  educationalLevels: [shortCycleTertiary, primary, earlyChildhood]
  spokenLanguages: [(Catalan-Valencian, b1), (Spanish-Castilian, b1)...]
  socioEconomicStati: [lowerClass, lowerMiddleClass]
  skillLevels: [beginner]
        
targetCommunity barcelonaCitizens [...]

adaptation DigitalDivide 
 description: "Training and Mediation Against the Digital Divide. Adaptations for bridging the gap of digital devices. Decidim has released..."
 targetCommunities: [nonDigitalSkilled] 
 
adaptation languageAdaptation
 description: "The cities where Decidim is deployed contain a broad variety of spoken languages. Along with a community of volunteers..."
 targetCommunities: [barcelonaCitizens]
 relatedTeams: [Translators]


\end{lstlisting}

Regarding the \emph{usage context}, Decidim has been \emph{adapted} to multiple linguistic and regional contexts, with translations into over 60 languages, accessibility features for people with disabilities, and initiatives to bridge the digital divide by integrating users with low digital skills. To support these users, the platform facilitates in-person events for public decisions and introduces ``managed users'', enabling participation through volunteer assistance. In Listing \ref{lst:decidim} we can see en excerpt of the usage context of the Decidim diversity card, where we have been able to capture the two adaptation \emph{DigitalDivide} and \emph{languageAdaptation} and to profile the target community of the \emph{DigitalDivide} adaptation defining the age range, location, educational and skill levels, and spoken languages.

They have also conducted several accessibility audits to adapt the software for people with disabilities. However, while various governments have conducted these audits, the results remain unpublished. During the interview, they commented that one of the reasons this information is not published is the lack of resources and clear guidance on how to publish these results, which are sometimes hard for final users to understand as one of the reasons. However, they agree that pointing to the results of audits would be useful for other governments willing to adopt Decidim in their participatory processes. \change{Similarly to the usability test, this shows a benefit of the standardization efforts like our proposal, which could guide on how to disclose accessibility information.}

The \emph{governance} of Decidim is based on public and democratic processes, and the board-elected members can be consulted online. Facing issues similar to those of the non-coding platform, the information needed to generate aggregated profiles is scarce as the users only need to be linked to a username. The funding is based on a mix of public administrations and crowd-sourcing platforms where individuals and organizations make donations. The Decidim governance profile can be seen in the online Decidim card.


\subsection{Besser: A low-code platform for smart software development}

Besser is a low-code platform that allows non-coding developers to develop domain-specific applications. It is maintained and developed by a community of \emph{participants}; for instance, the development team is composed of 15 developers from 9 nationalities, where French, English, and Spanish are the running languages, and the educational level is balanced between undergraduates and doctoral degree holders in computer science. Although this information is partially public on the project website, the interviews with the team have provided us with direct access to the development team and the necessary information.  Still, we needed to gather information on the geographic distribution, the languages used, and the education levels from the interviews with the project managers. Listing \ref{lst:besser} shows the different attributes captured in the card, \change{and Table \ref{tab:bessercom} shows an overview of the card's completeness.}

The project has an emerging community of \emph{non-coding roles} around those who report bugs. This community interacts around the project's open repository via issues and pull requests and less frequently through direct contacts or social networks. When asking for an aggregated profile, they faced challenges similar to those faced by Decidim in profiling this community, as the contributions are just associated with a username. Another similar point to Decidim is that they conducted a set of usability studies as part of the framework, but these have not been published. They realized the relevance of publishing this data in our conversation and shared with us the aggregated profiles that can be viewed in the complete example of the card.

\begin{table}[b!]
    \centering
\scriptsize
\renewcommand{\arraystretch}{1.3}
    \caption{Overview of the completeness of Besser's card.}
    \label{tab:bessercom}
    \begin{tabular}{|ll|l|}

       \hline
        \textbf{Dimensions} & \textbf{Team type} & \textbf{Profile completeness} \\ \hline   
        {\textbf{Participants}} & \textit{Developer Team} &  High, reported directly by development team \\ 
        \textbf{} & \textit{Non-coding contributors}  & Limited due to anonymity-first approach of OSS  \\ 
        \textbf{} & \textit{Testers}  & High, tests lead by software owners  \\ 
        \textbf{} & \textit{Reporters}  & Limited due to anonymity-first approach of OSS \\ 
         \hline
        {\textbf{Usage Context}} & \textit{Target Communities}  & Limited to the project's use cases \\ 
  
        \hline
      
        {\textbf{Governance}} & \textit{Boards}  &  High, principal investigator \\ 
       \textbf{} & \textit{Bodies}  &  High, public institutions  \\ %

        \hline

    \end{tabular}

\end{table}

\begin{lstlisting}[
label=lst:besser,
language=DSL,
float=t,
upquote=true,
caption= Excerpt of the development team part of the Besser's card.,
   ]
developmentTeam DevelopmentTeam
    description: "The Besser development team is composed of 15 developers, based in Luxembourg, from 11 different ethnic groups."
    startDate: 11-08-2022
    teamSize: 15
    ageRange: (25, 36)
    locations: ["Luxembourg Institute of Technology, Luxembourg"]
    workplaceType: presential
    ethnicities:
        ["Colombian", "Brasilian", "Argentinian", "French", "Spanish",
        "Pakistani", "Serbian", "Iranian", "Moroccan", "Italian"]
    genders: ["male 80%","female 20%","non-binary 0%"]
    religiousBeliefs: ["Christianism","Islam"]
    countries: [Luxembourg]
    educationalLevels: [masterEquivalent, doctorateEquivalent]
    spokenLanguages: [(English, c1)]
    averageTenure: 3.3
\end{lstlisting}

Regarding the \emph{usage context}, the adaptations are one of the key elements of Besser, as it is a low-code platform intended to help users develop domain-specific software. In that sense, the software presents a set of domain adaptations identifying a target user for each domain. For instance, they created an extension of Besser to allow public servants and researchers to create visualization panels for climate data; on the other hand, they adapted Besser for scholars, focusing on teachers and students in computer science, and an ongoing project involves adapting Besser to telecommunications engineers to model the deployment of 6G networks. Regarding \emph{governance}, Besser is a public-funded project funded by the FNR Pearl grant led by the Luxembourg Institute of Science and Technology. The principal investigator is the grant recipient and acts as a project director (unlike Decidim, which is governed by a board of directors). 




%
\change{\section{Discussion} \label{sec:challenges}}

\change{In this section, we discuss the benefits of the proposed Software Diversity Card for both practitioners and the research community, then we discuss the challenges that emerged during its application, and finally, we present the threats to the validity of our study. }

\change{For practitioners, the card represents an initial effort to recognize diversity as a critical dimension of software projects. In the applied use cases, the card enabled us to collect information that was not readily available in public documentation, despite its potential to enhance transparency and project value.  Additionally, the card provided guidance on which diversity-related information is important to share and offered a structured format for doing so. For example, in the Besser and Decidim projects, usability and accessibility tests were routinely conducted for internal purposes but rarely published, even though sharing these results could encourage adoption by institutions and public administrations, while also promoting transparency and accountability. Overall, the card serves as a practical tool for making diversity considerations more visible in software projects by practitioners. By highlighting key information and offering a structured approach to disclosure, it has the potential to enhance diversity-sharing practices and serve as a communication tool between practitioners, users, and public institutions. }

\change{For the research community, this initiative aims to stimulate greater attention to diversity in software development. Analogous to reproducibility badges, the Software Diversity Card could serve as a formal mechanism to encourage the inclusion of diversity considerations in submission guidelines and review procedures for scientific venues, particularly those publishing software-related work. By providing a structured way to report and reflect on diversity aspects, the card can help raise awareness among researchers and reviewers, fostering a culture that values and actively addresses diversity. We also recognize that diversity, like any social phenomenon, is dynamic and evolves alongside the societies in which software is developed and deployed. The card should therefore be viewed as an initial step rather than a definitive solution. Its adoption and refinement over time can support ongoing reflection, dialogue, and iterative improvements in how diversity is understood, measured, and promoted within the software development and research communities.}

\subsection{Challenges}

Several challenges have emerged during the application of the card. One of the main challenges encountered when creating the Decidim and Besser cards was the \textbf{anonymity} of the contributors. While anonymity is essential for reducing biases and protecting participants, it poses significant challenges for profiling contributors in open-source communities. For example, in Decidim, the “anonymity first” principle helps prevent discrimination against specific groups but complicates the analysis of non-coding contributor teams, such as translators or software reporters. This limitation restricts the ability to perform post-hoc analyses of community composition, making it more difficult to understand and support the broader user base effectively.

Closely related to anonymity is the challenge of handling \textbf{sensitive information} in the Software Diversity Card, including data on developers, contributors, and users, such as race or ethnicity, age, gender, and sexual orientation. While such information is typically aggregated, in smaller projects, even aggregate data may allow for individual identification, raising privacy concerns. Surveys and data collection processes must therefore include informed consent, clearly communicating what data is collected and how it will be used. Moreover, when population size or characteristics permit the inference of individual traits, this information should not be included in the card. \change{These issues underscore the importance of aligning the Software Diversity Card with current privacy regulations, such as GDPR. Ensuring compliance while capturing meaningful diversity-related insights remains an open challenge and represents a key direction for future research on the practical application and governance of the card.}

The software diversity card is a self-assessment report and, as such, it assumes goodwill on the part of the authors in terms of its \textbf{authenticity}. Nevertheless, the creators of a diversity card may have strong incentives to misrepresent the level of diversity in a software project, \emph{e.g.}, to attract more talent, to access funding from public organizations or specific communities, to gain publicity or public support, or to strengthen its participation in competitive calls. \change{One potential mechanism to alleviate this challenge is the inclusion of a third-party system to store and share card information, acting as a trust partner for external users.} It is important to emphasize that the purpose of the software diversity card is to improve community diversity practices in software development. Such a goal can only be achieved if the information presented in diversity cards is accurate.

One of the challenges during the creation of Decidim's card has been the \textbf{evolving nature of the development team}. The team comprises internal and external teams, which could easily change shortly. They pointed out that this situation makes disclosing the team's long-standing profile information difficult. The same problem could happen when new tests are done with users or the evolution of the reporting community. So, the inherent evolving nature of the project forces us to frequently reanalyze the card, which can be helpful to determine whether the project is improving or worsening in terms of diversity and inclusion.

Practitioners may encounter several \textbf{challenges when adopting} the Software Diversity Card. A primary concern is the additional effort required, which can be a significant barrier for small- and medium-sized projects, while larger projects may be better equipped to manage this complexity. This creates a potential risk of restricting competition, as smaller projects could be excluded from public tenders due to the administrative burden, which is both time-consuming and resource-intensive. To alleviate this issue, we developed a toolkit to support the use and creation of the card; however, addressing this challenge fully will require a mature toolkit ecosystem that evolves alongside the needs of the software development community. \change{An interesting direction for tool development is the integration of the Software Diversity Card with existing documentation assets, such as README files, codes of conduct, and governance documents. This integration could be achieved either by automatically generating or updating documentation from a valid card, or by populating the card by parsing existing documentation—leveraging, for example, the promising capabilities of Large Language Models in feature extraction from technical documents \cite{giner2023datadoc}.} Another important challenge is the lack of incentives for adopting diversity practices. Beyond the intrinsic benefits of promoting diversity, external mechanisms—such as reproducible badges awarded by public regulators, similar to those used in scientific publications—could serve as tangible incentives for organizations to implement and report on diversity measures.

\change{\subsection{Threats to validity} }

\change{A key limitation of this study is its focus on open-source software projects and their public documentation, which may restrict the generalizability of our findings to other contexts. While open-source communities prioritize transparency, they might have additional documentation not considered here. In contrast, commercial software platforms often limit access to documentation, which could skew the practices seen in open-source projects. Future research should expand the evaluation to include a wider range of software ecosystems and additional documentation sources.}

\change{A second threat arises from our use of large language models (LLMs), specifically GPT-4o-mini, for textual analysis and extraction of information. While LLMs have demonstrated strong performance in analyzing and summarizing text, they are prone to errors, omissions, and hallucinations, which could affect the validity of the extracted data. To mitigate this risk, we performed manual validation of the model outputs and cross-checked key findings with the project documentation and interviews. Nonetheless, the potential for subtle inaccuracies remains a consideration when interpreting the results.}

\change{Finally, the geographical context of the case studies introduces another potential threat to validity. Both the Besser and Decidim projects are based in Europe, and diversity-related aspects are highly sensitive to regional cultural and social norms. As such, our findings may not generalize to projects situated in other regions with different social, cultural, or regulatory contexts. Addressing this limitation will require future studies to examine software projects in more diverse geographical settings, which would provide a more comprehensive understanding of how diversity manifests across global software development practices.  }

\section{Conclusions and future work}
\label{sec:conclusions}

In this work, we examined the growing concerns in the research community regarding diversity in software projects, extending the focus beyond development teams to include non-coding roles, testers, and end users. Despite increasing awareness, our analysis of 1,000 popular open-source projects revealed a substantial lack of reported diversity information. To address this gap, we proposed the Software Diversity Card as an integrated approach for reporting diversity-related aspects across this broader definition of participation. To facilitate adoption, we also formalized a domain-specific language (DSL) and developed a supporting toolkit, providing a common framework for stakeholders to understand, communicate, and discuss diversity considerations effectively.

We validated the card through two real-world case studies, analyzing public documentation and conducting interviews with development teams. The findings illustrate both the benefits and challenges of using the card in practice. Among the benefits, the card enabled the collection of valuable information not publicly documented, such as internal usability test results, which could enhance project transparency and value. The challenges included issues related to anonymity and sensitivity, which limited the collection of nuanced participant data, the dynamic nature of development teams, which complicates capturing their full composition and evolution, and adoption barriers, particularly the additional effort required for small- and medium-sized projects. These results highlight the potential of the Software Diversity Card as a tool for promoting awareness and reporting of diversity, while also indicating areas for further refinement and support to ensure practical applicability across diverse software projects.

In the remainder of this section, we outline several directions for future research and development in reporting diversity-related aspects of software projects: 

\smallskip
\noindent \textbf{Integrating diversity tracking to specific development process tasks:} Right now, we model the diversity of a project as a whole, but we could follow a more fine-grained approach where diversity is also tracked at the task level for deeper insights. For example, we could analyze whether a project has a diverse team, but some tasks are only targeted by some team members, resulting in a less diverse approach for them (such as the project managers being mainly men). \change{In addition, it would be interesting to account for the level of involvement of the individuals, as different levels of involvement could also raise bias and insights into diversity in a team.}  To generate these insights, it is essential to integrate Software Diversity Cards with software process definition tools and languages, such as SPEM\footnote{SPEM specification homepage: \url{https://www.omg.org/spec/SPEM/2.0/About-SPEM}}.

\smallskip
\noindent \textbf{Formally verifying software compliance with diversity requirements:} Public administrations can utilize the Software Diversity Card to assess which software artefacts meet their mandatory diversity requirements. Since the Software Diversity Card is precisely specified, query languages like OCL \cite{CabotG12} can be employed to assist and automate this compliance-checking process.

\smallskip
\noindent \textbf{Enhancing the diversity card with environmental information:} Alongside diversity, there is a growing concern regarding the environmental aspects of software development. For instance, in the machine learning (ML) field, transparency about the environmental impact of AI training has become a significant topic. Initiatives like MyGreenLabel\footnote{MyGreenLabel homepage: \url{https://act.mygreenlab.org}} provide insights into the carbon footprint of trained ML models. Incorporating this environmental information into the Software Diversity Card framework could further contribute to the creation of ``Responsible Software.'' Exploring methods to integrate and balance these two types of information would be a fruitful avenue for research.

\smallskip
\noindent \textbf{Bridging the gap between diversity and responsible AI documentation:} Machine learning applications are software systems often integrated into larger traditional software frameworks. Therefore, documentation assets for ML, such as Data Cards \cite{pushkarna2022data}, can be viewed as a subset of the software diversity card. By documenting diversity factors related to the development team and beta testing population, we can also capture diversity aspects involved in designing and training ML models within these systems. In addition, several efforts have been made to translate these RAI data documentation frameworks to machine-readable formats, mainly to enhance their processing and discovery. Building mapping and translation between the software diversity card to data machine-readable vocabularies, such as \mbox{DescribeML \cite{giner2023domain}} and Croissant \cite{akhtar2024croissant} could also enhance the discoverability of diversity aspects on the web.

\smallskip
\noindent \textbf{AI agents as biased actors:} AI agents are becoming increasingly important in software projects. It is essential to consider the biases present in these models, as they can impact the diversity of the entire software project. Several works in the software engineering (SE) and machine learning (ML) communities have started to identify and address these biases. Looking ahead, there is a need to incorporate this information into the software diversity card.

\smallskip  
\noindent \textbf{Empirical evaluation of the software diversity card's impact:}  
While the software diversity card provides a structured approach to documenting diversity, its future impact on software projects will need to be systematically assessed. Future work should explore its influence on inclusion through case studies in open-source and private projects, analyzing metrics such as participation of underrepresented groups, diversity perceptions in OSS communities, and improvements in accessibility. These studies should point in two directions, whether the diversity card approach is adopted by OSS projects (for instance, by performing longitudinal studies on \textsc{GitHub}) and other institutions, like public administration in the call for tenders, and whether the projects that adopt it see some positive impact from it.

\section*{Acknowledgements}
Joan Giner-Miguelez acknowledges his AI4S fellowship within the “Generación D” initiative by Red.es, Ministerio para la Transformación Digital y de la Función Pública, for talent attraction (C005/24-ED CV1), funded by NextGenerationEU through PRTR. 
This work is part of the projects TED2021-130331B-I00 and PID2023-147592OB-I00, and the research network RED2022-134647-T, all funded by MCIN/ AEI/10.13039/501100011033; and the Luxembourg National Research Fund (FNR) PEARL program, grant agreement 16544475.

\label{}







\bibliographystyle{elsarticle-num} 
\bibliography{sample}



\end{document}